# Study on the Electronic Structure and Stability of Some Endohedral Fullerenes - $RE_3N@C_{80}$ (RE = Sc, Y, La) by PM7

Kye-Ryong Sin[1], Sun-Gyong Ko[1], Kwang-Yong Ri[1], and Song-Jin Im[2]
[1]Department of Chemistry and [2]Department of Physics,
**Kim Il Sung** University, Daesong district,
Pyongyang, DPR Korea
E-mail: ryongnam9@yahoo.com

**Abstract**: In this paper, we investigated the electronic structure of some endohedral fullerenes - $RE_3N@C_{80}$ ( RE = Sc, Y, La) and the stability of their benzyne – adducts by using PM7, the updated version of the semi-empirical Hartree-Fock method. Introduction of $RE_3N$ into $C_{80}$ stabilized the fullerene with transfer of electrons from $RE_3N$ to $C_{80}$. The results showed that the stable addition of benzyne on the fullerene may take place on the different positions of its surface depending on whether it contains $RE_3N$ or not.

## 1. Introduction

Recently, many researches have been carried out for synthesis of the endohedral fullerenes, the so-called "cluster-fullerene", containing metal atoms or clusters therein and for their application in manufacture of the useful nano-materials such as molecular devices and molecular medicines[1]. From the first preparation of $La@C_{82}$ in 1991, there made lots of advances in synthesis of the endohedral fullerenes with metal atoms or their nitride.

Since the cluster fullerenes have peculiar properties useful in electronics, magnetics and photonics, they could enjoy the wider prosperity in many applications such as molecular electronics and bio-technology.[2,3] The recent reports showed the possibility of control of the stability, reactivity, solubility and function of the endohedral fullerenes by adding some organic compound on their surfaces.[7,10]

In this paper, PM7, one of the updated version of the semi-empirical Hartree-Fock methods, was applied in the theoretical study on the electronic structure and stability of $RE_3N@C_{80}$ (RE = Sc, Y, La) and their benzyne (BZ) adduct - $RE_3N@C_{80}$-BZ. There have been some reports on DFT (Density Functional Theory) study on the cluster –fullerenes, but no one on PM7 application for it.

## 2. Computational Models and Method

For calculating the electronic structure of $RE_3N@C_{80}$ (RE = Sc, Y, La), two geometric isomers of $C_{80}$ ($C_{80}$-$I_h$ and $C_{80}$-$D5_h$) were chosen as the fullerene cage – $C_{80}$, where $I_h$ and $D_{5h}$ show the geometric symmetry of the fullerene cages. (Figure 1)

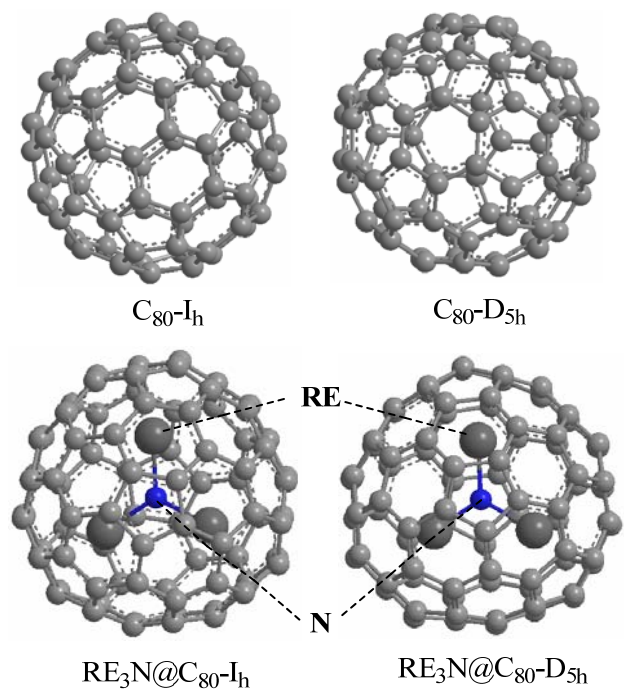

Figure 1. Models for $RE_3N@C_{80}$

Figure 2 shows the models for evaluating the effects of rare earth nitride – "mixed $RE_3N$" containing two or three kinds of RE (Sc, Y, La) on the electronic structure of $C_{80}$-$I_h$.

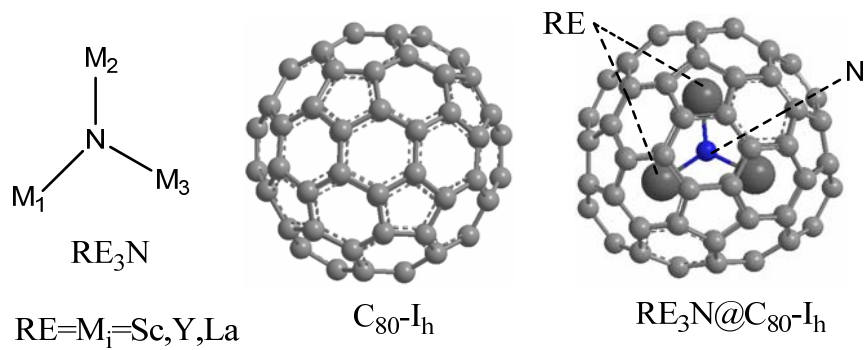

Figure 2. Models for mixed $RE_3N@C_{80}$-$I_h$

To locate the reaction site for the stable products of BZ addition to $C_{80}$-$I_h$ or $RE_3N@C_{80}$-$I_h$, two different reaction positions ([5,6] position and [6,6] position) were considered, where [5,6] means C=C bond belonging to both of the neighbored 5-membered carbon cycle and 6-membered carbon cycle simultaneously and [6,6] means C=C bond belonging to both of two neighbored 6-membered carbon cycles. (Figure 3)

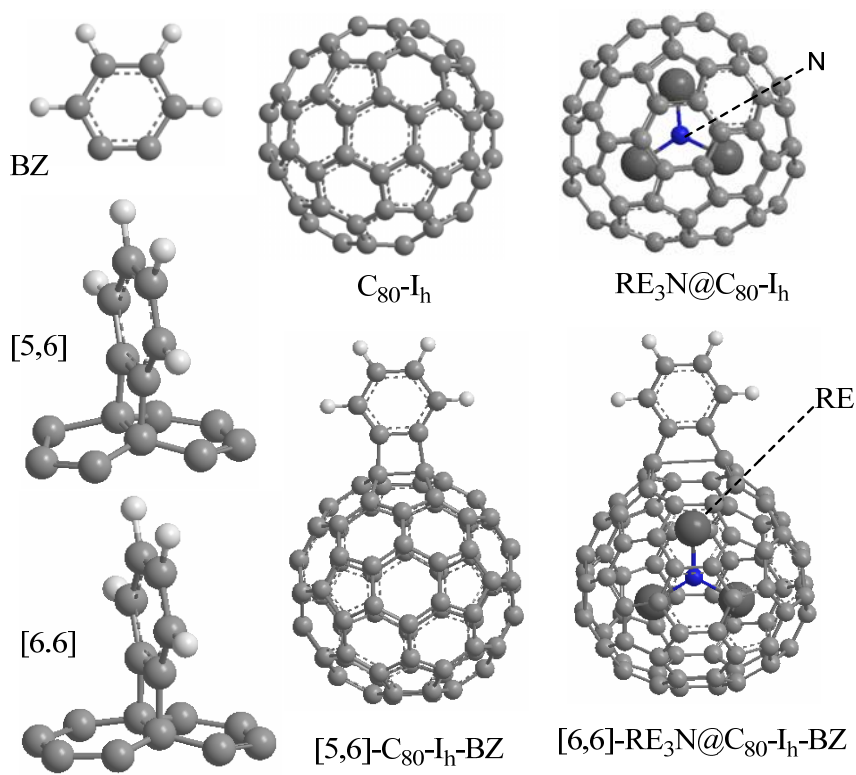

Figure 3.　Models for BZ addition on $C_{80}$-$I_h$

The geometric and electronic structures of the models were calculated by PM7 from **MOPAC2012**, the latest version of the semi-empirical MO software package, that is well-known as one of the most efficient semi-empirical Hartree-Fock methods with the enhanced accuracy for a wide range of molecules, complexes, polymers, and crystals.[5] Especially, it offers good parameter set for the calculation of most of the elements on the periodic table including rare earth elements.

The first step of calculation was the geometry optimization by EF (Eigenvector-Following) routine, which was followed by the single-point MO and energy calculation.

Stability of the models was evaluated by $\Delta E_t$, the difference of total energy ($E_t$) between the resulting model and the starting one.

The obtained results were compared with the previous ones from the experiments or from those of the accurate DFT calculations wherever they were available.

### 3. Results and Discussion

**1) The electronic structures of $C_{80}$ and $RE_3N@C_{80}$**

The previous experimental researches showed that it was very difficult to obtain $C_{80}$-$I_h$ and $C_{80}$-$D_{5h}$ in quantity because of their low stability, but the endohedral $C_{80}$ with rare earth metal atoms or their nitride had the increased stability that made it possible to prepare those endohedral fullerenes quantitatively.[6]

It can be seen from the optimized geometric structures of RE$_3$N@C$_{80}$ in Figure 4 that RE$_3$N (RE = Sc, La) has the plane form in both of C$_{80}$-I$_h$ and C$_{80}$-D$_{5h}$, but Y$_3$N has the pyramid form, which resembles the previous XRD measurement of Gd$_3$N@C$_{80}$-I$_h$. [9]

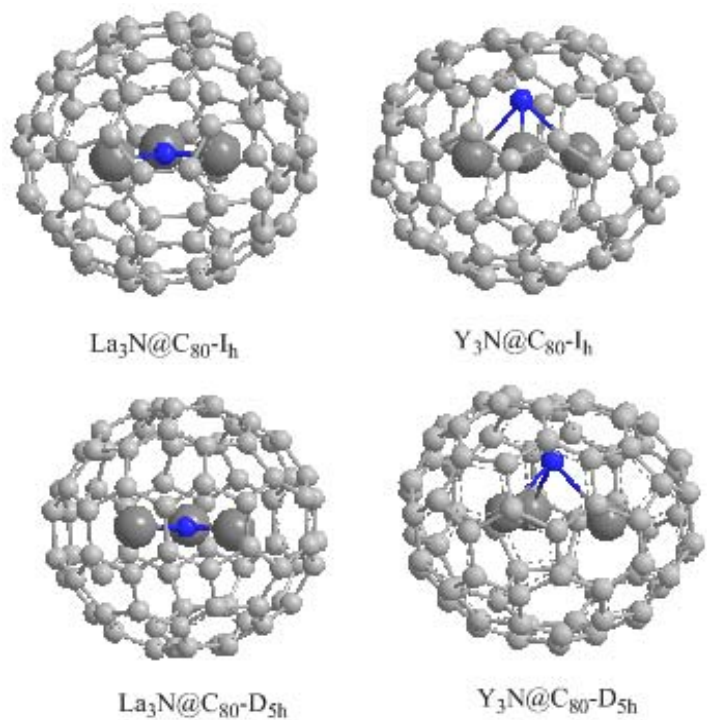

Figure 4. The optimized structure of some RE$_3$N@C$_{80}$

In the optimized RE$_3$N@C$_{80}$ models the bond lengths of Sc-N and La-N were shortened from those of free RE$_3$N, but that of Y-N increased. (Table 1) Similar results were obtained by DFT calculation on Y$_3$N@C$_{78}$-D$_{3h}$.[4] Bond lengths of Sc-N and Sc-C were also well compared with XRD measurements on Sc$_3$N@C$_{80}$-I$_h$.[7]

Table 1.     Bond lengths (nm) in free RE$_3$N and RE$_3$N@C$_{80}$

| fullerene cage | C$_{80}$-I$_h$ | | | C$_{80}$-D$_{5h}$ | | |
|---|---|---|---|---|---|---|
| RE | Sc | Y | La | Sc | Y | La |
| RE-N in free RE$_3$N | 0.2639 | 0.2262 | 0.2903 | 0.2639 | 0.2262 | 0.2903 |
| RE-N in RE$_3$N@C$_{80}$ | 0.1938 | 0.2463 | 0.1809 | 0.1903 | 0.2466 | 0.1808 |
| RE-C ( the nearest) | 0.2488 | 0.2727 | 0.2626 | 0.2467 | 0.2704 | 0.2591 |

From the electronic structures calculated from the optimized geometry, it was found out that the positive charge of RE atoms was increased and the negative charge of N atom decreased in the cluster fullerene compared with those in free RE$_3$N, which shows more portion of electrons of RE transferred to fullerene cage than those to N. In Table 2, C$^-$ means the carbon atom with the maximum negative charge among those of the fullerene cage.

Table 2.　　The electronic structures of $C_{80}$ and $RE_3N@C_{80}$

| fullerene | RE | μ (D) | atomic charge (e) | | |
|---|---|---|---|---|---|
| | | | RE | N | C$^-$ |
| $C_{80}$-$I_h$ | - | 0.016 | - | - | -0.064 |
| | Sc | 0.133 | 2.031 | -1.901 | -0.241 |
| | Y | 2.885 | 1.560 | -1.365 | -0.171 |
| | La | 0.032 | 1.989 | -1.591 | -0.130 |
| $C_{80}$-$D_{5h}$ | - | 6.492 | - | - | -0.221 |
| | Sc | 0.565 | 2.041 | -1.888 | -0.236 |
| | Y | 2.721 | 1.154 | -1.361 | -0.184 |
| | La | 0.020 | 1.986 | -1.594 | -0.132 |

Changes of the electronic structure of the cluster fullerenes by encapsulation of RE$_3$N were different according to their geometric symmetry. In RE$_3$N@C$_{80}$-I$_h$, its dipole moment (μ) and the maximum negative atomic charge (Q$_{C^-}$) among 80 carbon atoms increased, but in RE$_3$N@C$_{80}$-D$_{5h}$, μ was decreased and Q$_{C^-}$ decreased or unchanged. Both of Y$_3$N@C$_{80}$-I$_h$ and Y$_3$N@C$_{80}$-D$_{5h}$ had much larger dipole moment than others because in these cluster-fullerenes N atom was located over the Y-Y-Y plane, no forming the plane structure like others, but forming the pyramid structure like in the previous DFT study on Y$_3$N@C$_{78}$-D$_{3h}$.[8]

In case of the empty C$_{80}$ without RE$_3$N, C$_{80}$-D$_{5h}$ was a bit more stable than C$_{80}$-I$_h$. Inserting RE$_3$N (RE = Sc, La) into C$_{80}$ made the fullerene more stable, especially in C$_{80}$-I$_h$. But Y$_3$N made both of C$_{80}$ isomers ( I$_h$ and D$_{5h}$ ) unstable, more apparently in C$_{80}$-D$_{5h}$. (Table 3)

Table 3.　　　　Total energy (eV) of RE$_3$N@C$_{80}$

| RE | C$_{80}$-I$_h$ | | | C$_{80}$-D$_{5h}$ | | |
|---|---|---|---|---|---|---|
| | starting state | resulting state | ΔE$_t$ | starting state | resulting state | ΔE$_t$ |
| Sc | -10104.9 | -10110.8 | -5.9 | -10106.4 | -10106.5 | -0.1 |
| Y | -10091.5 | -10081.6 | 9.9 | -10093.0 | -10080.3 | 12.7 |
| La | -10103.9 | -10115.2 | -11.3 | -10105.4 | -10114.0 | -8.6 |

### 2) Stability of the mixed RE$_3$N@C$_{80}$-I$_h$

To simplify the problem, all of the fullerene cages discussed below was fixed as C$_{80}$-I$_h$.

Figure 5 shows the optimized geometric structures of the mixed RE$_3$N@C$_{80}$, where most of the mixed RE$_3$N had the plane form, but the mixed RE$_3$N containing two Y atoms, Y$_2$ScN@C$_{80}$ and Y$_2$LaN@C$_{80}$, had the pyramid form, which are similar to the previous result from DFT calculation.[8]

In table 4, $\triangle H_f$ (heat of formation) and $\Delta E_t$ of the mixed $RE_3N@C_{80}$ shows that the more portion of electrons transferred from the mixed $RE_3N$ to $C_{80}$, the more stability it obtained. It was already known that charge transfer ($\triangle Q$) from the encapsulated cluster to the fullerene cage is the main reason of the stabilization of the cluster-fullerene. [4]

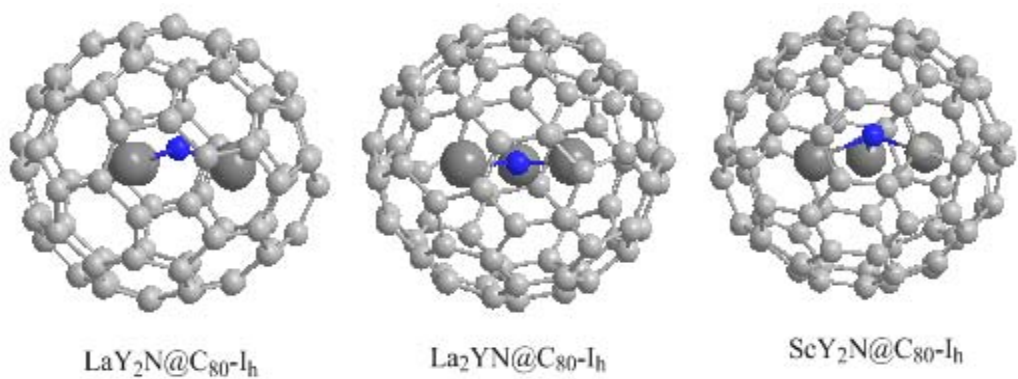

Figure 5. The optimized structure of the mixed $RE_3N@ C_{80}$- $I_h$

Table 4.    The electronic structure of mixed $RE_3N@ C_{80}$- $I_h$

| $RE_3N$ | $\triangle H_f$ (kJ·mol$^{-1}$) | $\Delta E_t$ (eV) | $\varepsilon$ (eV) | | $\triangle Q$ (e) |
|---|---|---|---|---|---|
| | | | HOMO | LUMO | $RE_3N \Rightarrow C_{80}$ |
| $Sc_2YN$ | 4544 | -1.16 | -9.575 | -4.512 | -4.097 |
| $Sc_2LaN$ | 3473 | -6.00 | -9.740 | -4.955 | -4.213 |
| $Y_2ScN$ | 5269 | 5.39 | -9.281 | -4.881 | -3.989 |
| $Y_2LaN$ | 4684 | 4.55 | -9.318 | -5.075 | -4.090 |
| $La_2ScN$ | 2961 | -7.50 | -9.676 | -5.053 | -4.274 |
| $La_2YN$ | 3456 | -2.58 | -9.497 | -4.478 | -4.242 |
| $ScYLaN$ | 4017 | -1.05 | -9.495 | -4.717 | -4.152 |

From the above results, it was found out that more portion of La and Sc in the mixed $RE_3N$ offers higher stability to their mixed $RE_3N@C_{80}$, but more Y makes it unstable.

### 3) Addition of benzyne to $RE_3N@ C_{80}$- $I_h$

Among [5,6]-addition products (Table 5, Figure 6), $C_{80}$-$I_h$-BZ has the largest surface and the smallest volume, and all of $RE_3N@C_{80}$-$I_h$-BZ have smaller surface and larger volume. $\Delta E_t$ shows that $RE_3N@C_{80}$-$I_h$-BZ is more stable than $C_{80}$-$I_h$-BZ.

Table 5.    Geometric and electronic structure of [5,6]-adducts

| adduct | $C_{80}$-BZ | $Sc_3N@C_{80}$-BZ | $Y_3N@C_{80}$-BZ | $La_3N@C_{80}$-BZ |
|---|---|---|---|---|
| surface (nm$^2$) | 5.736 | 5.229 | 5.303 | 5.233 |
| volume ( nm$^3$) | 0.916 | 0.971 | 0.985 | 0.966 |
| μ (D) | 6.657 | 6.383 | 6.253 | 6.501 |
| $E_t$ (eV) | -10590.8 | -10903.3 | -10875.4 | -10907.5 |
| $\Delta E_t$ (eV) | -4.6 | -6.0 | -7.3 | -5.8 |

| $\varepsilon_{HOMO}$ (eV) | -8.619 | -9.556 | -8.802 | -9.301 |
|---|---|---|---|---|
| $\varepsilon_{LUMO}$ (eV) | -4.263 | -4.581 | -3.984 | -4.533 |

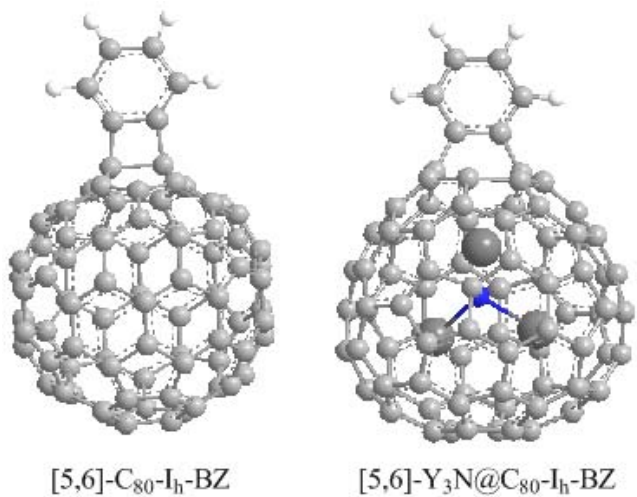

Figure 6. The optimized structure of [5,6]-adducts

To compare the relative stability of [6,6]-adducts with that of [5,6]-adducts, $\Delta E_t(66{:}56)$ was calculated as the $E_t$ difference between [6,6]-adduct and [5,6]-adduct from the same fullerene or from the same cluster fullerene. In case of the empty $C_{80}$-$I_h$-BZ, $\Delta E_t(66{:}56)$ shows [5,6]-adduct is much more stable than [6,6]-adduct. But in case of $RE_3N@C_{80}$-$I_h$-BZ, [6,6] adduct is a bit more stable than [5,6] adduct. (Figure 7, Table 6)

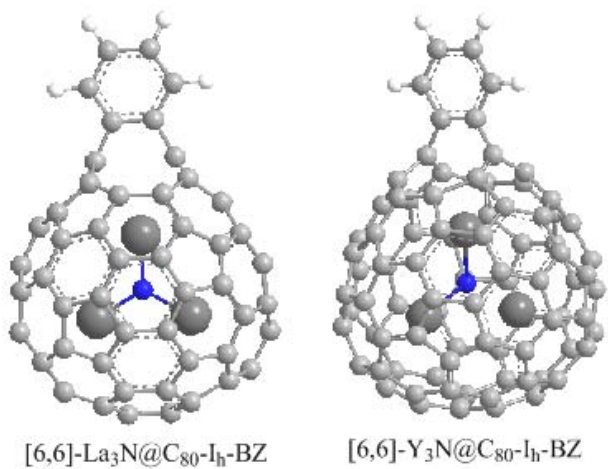

Figure 7. The optimized structure of [6,6]-adducts

Table 6.  Geometric and electronic structure of [6,6]-adducts

| adduct | $C_{80}$-BZ | $Sc_3N@C_{80}$-BZ | $Y_3N@C_{80}$-BZ | $La_3N@C_{80}$-BZ |
|---|---|---|---|---|
| Surface (nm$^2$) | 6.035 | 5.251 | 5.299 | 5.241 |
| Volume (nm$^3$) | 0.985 | 0.970 | 0.987 | 0.968 |
| μ (D) | 3.092 | 7.568 | 6.403 | 7.097 |
| $E_t$ (eV) | -10569.7 | -10903.9 | -10875.8 | -10908.0 |
| $\Delta E_t(66{:}56)$ (eV) | 21.1 | -0.6 | -0.4 | -0.5 |
| $\varepsilon_{HOMO}$ (eV) | -8.179 | -9.405 | -8.732 | -9.244 |
| $\varepsilon_{LUMO}$ (eV) | -3.507 | -4.561 | -4.238 | -4.533 |

These results showed that benzyne addition may take place on the different site of the fullerene surface depending on whether it has RE$_3$N in it or not. The empty C$_{80}$ without RE$_3$N prefers [5,6]-addition to [6,6]-addition, but RE$_3$N@C$_{80}$ seems to like [6,6]-addition with no remarkable dislike to [5,6]-addition. The kind of rare earth elements may also make some changes to the stability and reactivity of the endohedral fullerenes where they are wrapped.

## 4. Conclusion

Rare earth atoms in RE$_3$N@C$_{80}$ are the strong electron donors to the fullerene cage. RE$_3$N (RE=Sc, La) makes the fullerene more stable, but Y$_3$N seems to be unfavorable for it.

Benzyne addition to the fullerene may take place on the different sites of its surface depending on whether it contains RE$_3$N or not.

## Acknowledgement

The authors thank Stewart Computational Chemistry for its efficient **MOPAC2012**.

## References

[1] Silvia Osuna, Marcel Swart, and Miquel Sola, *Phys. Chem. Chem. Phys.*, **2011**, *13*, 3585–3603.

[2] Christopher Scott Berger, John W. Marks, Robert D. Bolskar, Michael G. Rosenblum, and Lon J. Wilson, *Translational Oncology*, **2011**, *4*, 350–354.

[3] Jie Meng, Dong-liang Wang, Paul C. Wang, Lee Jia, Chunying Chen, and Xing-Jie Liang, *J. Nanosci. Nanotechnol.*, **2010**, *10*, 1-7.

[4] Brandon Q. Mercado, Melissa A. Stuart, Mary A. Mackey, Jane E. Pickens, Bridget S. Confait, Steven Stevenson, Michael L. Easterling, Ramon Valencia, Antonio Rodriguez-Fortea, Josep M. Poblet, Marilyn M. Olmstead, and Alan L. Balch, *J. Am. Chem. Soc.*, **2010**, *132*, 12098-12105.

[5] James J. P. Stewart, *J. Mol. Model.*, **2013**, *19*, 1–32.

[6] Lothar Dunsch, Shangfeng Yang, Lin Zhang, Anna Svitova, Steffen Oswald, and Alexey A. Popov, *J. Am. Chem. Soc.*, **2010**, *132*, 5413-5421.

[7] Fang-Fang Li, Julio R. Pinzon, Brandon Q. Mercado, Marilyn M. Olmstead, Alan L. Balch, and Luis Echegoyen, *J. Am. Chem. Soc.*, **2011**, *133*, 1563–1571.

[8] Silvia Osuna, Marcel Swart, and Miquel Sola, *J. Am. Chem. Soc.*, **2009**, *131*, 129–139.

[9] Stevenson, S., Phillips, J. P., Reid, J. E., Olmstead, M. M., Rath, S. P., and Balch, A.L., *Chem. Commun.*, **2004**, 2814-2815.

[10] Ting Cai, Liaosa Xu, Chunying Shu, Hunter A. Champion, Jonathan E. Reid, Clemens Anklin, Mark R. Anderson, Harry W. Gibson, and Harry C. Dorn, *J. Am. Chem. Soc.*, **2008**, *130*, 2136-2137.

[11] Michio Yamada, Mari Minowa, Satoru Sato, Masahiro Kako, Zdenek Slanina, Naomi Mizorogi, Takahiro Tsuchiya, Yutaka Maeda, Shigeru Nagase, and Takeshi Akasaka, *J. Am. Chem. Soc.*, **2010**, *132*, 17953–17960.